\documentclass[lettersize,journal]{IEEEtran}
\usepackage{amsmath,amsfonts}
\usepackage{array}
\usepackage{tikz}
\usepackage{amsmath,mathtools}
\usepackage{amssymb}
\usepackage{enumitem}
\usepackage{booktabs}
\usepackage{graphicx}
\usepackage{amsmath}
\usepackage[table]{xcolor}

\usepackage{algorithmic}
\usepackage{subfig}
\usetikzlibrary{shadows, positioning, arrows.meta, calc, chains, fit, backgrounds, fadings}
\usepackage{booktabs}
 \usepackage{tabularx}
 \usepackage{array}

\usepackage{algorithm}
\usepackage{boxedminipage}
\usepackage{fancyhdr}
\usetikzlibrary{arrows.meta, positioning, calc, shadows, backgrounds, fit}
\usepackage{multirow}
\usetikzlibrary{arrows.meta, chains, shapes.symbols, shadows}

\usepackage{mathptmx}
\usepackage{array}
\usepackage{textcomp}
\usetikzlibrary{positioning}
\usetikzlibrary{arrows.meta}
\usetikzlibrary{calc}
\usetikzlibrary{shapes.geometric}
\usetikzlibrary{fit}
\usepackage{stfloats}
\usepackage{csquotes} 
\usepackage{url}
\usepackage{verbatim}
\usepackage{cite}
\usepackage{eqnarray}
\usepackage{stfloats}
\usepackage{url}
\usepackage{verbatim}
\usepackage{graphicx}
\def\BibTeX{{\rm B\kern-.05em{\sc i\kern-.025em b}\kern-.08em
    T\kern-.1667em\lower.7ex\hbox{E}\kern-.125emX}}
\usepackage{balance}
\usepackage{pifont}

\begin{document}
\title{Fluid Antenna Networks Beyond Beamforming: An AI-Native Control Paradigm for 6G}

\author{
    Ian F. Akyildiz, \textit{Life Fellow, IEEE} and Tuğçe Bilen, \textit{Member, IEEE}
    \thanks{Ian F. Akyildiz is with Truva Inc., Alpharetta, GA 30022, USA (e-mail: ian@truvainc.com).}%
    \thanks{Tuğçe Bilen is with the Department of Artificial Intelligence and Data Engineering, Istanbul Technical University, Istanbul, Turkey (e-mail: bilent@itu.edu.tr).}%
}


\maketitle
\begin{abstract}
Fluid Antenna Systems (FAS) introduce a new degree of freedom for wireless networks by enabling the physical antenna position to adapt dynamically to changing radio conditions. While existing studies primarily emphasize physical-layer gains, their broader implications for network operation remain largely unexplored. Once antennas become reconfigurable entities, antenna positioning naturally becomes part of the network control problem rather than a standalone optimization task.
This article presents an AI-native perspective on fluid antenna networks for future 6G systems. Instead of treating antenna repositioning as an isolated operation, we consider a closed-loop control architecture in which antenna adaptation is jointly managed with conventional radio resource management (RRM) functions. Within this framework, real-time network observations are translated into coordinated antenna and resource configuration decisions that respond to user mobility, traffic demand, and evolving interference conditions.
To address the complexity of multi-cell environments, we explore a multi-agent reinforcement learning (MARL) approach that enables distributed and adaptive control across base stations. Illustrative results show that intelligent antenna adaptation yields consistent performance gains, particularly at the cell edge, while also reducing inter-cell interference. These findings suggest that the true potential of fluid antenna systems lies not only in reconfigurable hardware, but in intelligent network control architectures that can effectively exploit this additional spatial degree of freedom.
\end{abstract}


\thispagestyle{fancy}

\pagestyle{fancy}
\fancyhf{}
\fancyhead[C]{\scriptsize Submitted to IEEE Wireless Communications Magazine, 2026}

\section{Introduction}

Wireless communication systems have steadily evolved toward greater flexibility and programmability. Early cellular networks relied on largely static infrastructures, where antennas, transmission strategies, and propagation environments offered limited adaptability during operation. Over time, however, wireless technologies have introduced increasing levels of reconfigurability. Massive multiple-input multiple-output (MIMO) systems enabled programmable spatial processing, while reconfigurable intelligent surfaces (RIS) allowed direct manipulation of the propagation environment \cite{11153881}. Together, these advances reflect a broader shift: physical components that were once static are becoming controllable and adaptive.

\begin{figure*}[t]
\centering
\includegraphics[width=0.7\textwidth]{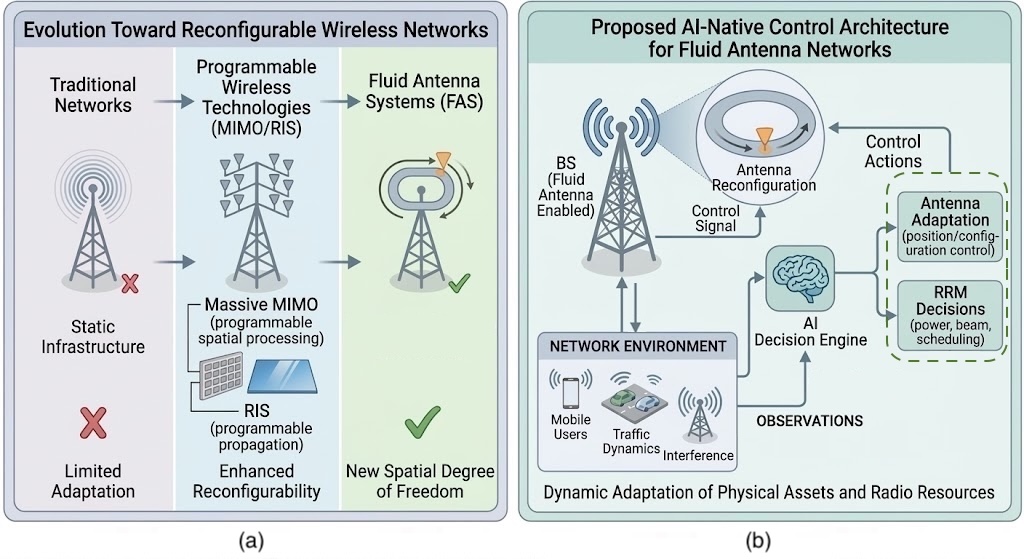}
\caption{(a) Evolution toward reconfigurable wireless networks, from static infrastructures to programmable wireless technologies and fluid antenna systems (FAS). (b) Proposed AI-native closed-loop control architecture for fluid antenna networks, where network observations drive intelligent antenna adaptation and radio resource management decisions.}
\label{model}
\end{figure*}

Fluid antenna systems (FAS) represent a further step in this evolution. Unlike conventional antennas with fixed positions and radiation characteristics, fluid antennas allow their radiating elements to move within a confined region \cite{11175437}. By adapting their position or configuration to radio conditions, they can dynamically reshape the effective channel experienced by users. In contrast to MIMO and RIS, which optimize transmission over a given environment, FAS enables direct physical control of the antenna itself, allowing the network to influence the channel realization. This introduces a new spatial degree of freedom, as illustrated in Fig.~\ref{model}, extending adaptation beyond signal processing to the physical antenna interface.

While FAS is often viewed as a physical-layer enhancement that improves channel quality and exploits spatial diversity, its implications extend beyond the physical layer. In large-scale cellular systems, antenna configuration becomes intertwined with user scheduling, beam management, power allocation, and inter-cell interference coordination. Since wireless networks continuously adapt to user mobility, traffic demand, and channel variations, dynamically adjustable antennas naturally introduce an additional control variable. A base station may reposition its antenna, adjust beamforming strategies, or adapt resource allocation in response to network conditions, leading to a tight coupling between antenna adaptation and radio resource management.

This perspective suggests that FAS should be viewed not merely as an antenna design paradigm, but as a new dimension of network control. Antennas evolve from fixed communication interfaces into dynamic, state-dependent network assets. The key challenge is therefore to determine how and when antenna configurations should be adapted under dynamic conditions. Artificial intelligence (AI) provides a natural approach to this problem, as learning-based methods can handle complex, time-varying environments where analytical modeling becomes difficult. Fluid antenna networks exhibit exactly such characteristics, with large configuration spaces and performance that depends on user distribution, mobility, and interference dynamics.

Motivated by these observations, this article explores an AI-native perspective on fluid antenna networks. Rather than focusing solely on antenna-level optimization, we examine how fluid antennas can be integrated into intelligent network control architectures that adapt to changing network conditions. In particular, antenna reconfiguration is embedded within a closed-loop decision process that jointly manages antenna adaptation and conventional radio resource allocation. We introduce an AI-driven control framework for fluid antenna networks and discuss how learning-based approaches enable distributed and scalable decision-making across multiple base stations. As a representative example, we consider a multi-agent reinforcement learning (MARL) approach that allows base stations to coordinate antenna positioning, beamforming, and resource allocation under dynamic conditions. We further present illustrative simulation results demonstrating how adaptive antenna configurations can improve network throughput, enhance cell-edge performance, and mitigate inter-cell interference compared with conventional fixed-antenna systems. Finally, we discuss practical implementation challenges and outline several research directions for fluid antenna networks in future 6G systems.

\section{Reconfigurable Wireless Systems: From PHY Optimization to Network Control}

Reconfigurable wireless technologies have been extensively studied to improve the adaptability and performance of communication systems. MIMO enables programmable spatial processing through large antenna arrays \cite{Marzetta2010MIMO}, while RIS allows dynamic manipulation of the propagation environment via controllable reflections \cite{Basar2019RIS, Wu2020RIS}. These approaches significantly enhance spectral efficiency and coverage. However, they fundamentally operate under fixed antenna infrastructures and do not treat antenna position as a controllable variable.

Building on this trend, FAS have recently emerged as a novel paradigm that enables dynamic reconfiguration of antenna positions, introducing a new spatial degree of freedom for wireless communications. Existing studies primarily focus on physical-layer aspects, including channel gain optimization, spatial diversity enhancement, and interference mitigation. For example, relay-assisted and RIS-integrated FAS architectures have demonstrated notable improvements in spectral efficiency and reliability \cite{Wang2024FAS_DRL, Xu2025FAR, Chen2025FAS_RIS, Salem2025FRIS}. However, these works are largely limited to link-level or single-cell scenarios and do not explicitly address network-level control and multi-cell interactions.

AI techniques have also been widely applied to radio resource management (RRM) in wireless networks. Reinforcement learning (RL) and MARL have been used for user scheduling, power control, and interference coordination in dynamic environments \cite{Luong2019RL, Zhang2021MARL, Challita2019RRM}. These approaches provide scalable solutions for high-dimensional and time-varying systems. However, they typically assume fixed antenna configurations and do not consider antenna reconfigurability as part of the control space.

Taken together, these research directions reveal a clear disconnect in the literature. Reconfigurable wireless technologies, including fluid antenna systems, have been predominantly investigated from a physical-layer perspective, focusing on channel enhancement, diversity gains, and interference mitigation, often in isolated or single-cell settings. In parallel, AI-driven approaches have been extensively developed for network-level optimization, particularly for radio resource management, but typically under the assumption of fixed antenna infrastructures.

As a result, the interplay between antenna reconfigurability and network-level control decisions remains insufficiently explored, especially in multi-cell and dynamic environments. While recent efforts have begun to incorporate learning-based techniques into reconfigurable systems, antenna configuration is still generally treated as a standalone optimization variable rather than an integral component of the network control space.

In contrast, this work adopts a unified control perspective in which antenna configuration is explicitly embedded into the network decision process. Specifically, antenna adaptation is jointly considered with conventional radio resource management functions within a learning-driven framework, enabling coordinated decision-making across multiple cells. By treating antenna states as part of both the system state and the control action space, the proposed approach bridges the gap between physical-layer reconfigurability and network-level intelligence, positioning fluid antennas not merely as a link-level enhancement but as a fundamental control dimension for future 6G networks.

\section{Fluid Antenna Networks as a New Control Dimension}
Fluid antenna systems introduce a new degree of freedom by enabling dynamic adaptation of antenna configurations. While often studied for channel enhancement and diversity gains, their impact extends beyond the physical layer. In multi-user and multi-cell environments, antenna configuration decisions interact with scheduling, beamforming, and interference coordination, thereby introducing an additional control dimension within the radio access network.

A fluid antenna-enabled base station can be viewed as a dynamic system whose operational state evolves over time. At any given time $t$, the antenna configuration of base station $b$ can be represented as $A_b(t) = \big(p_b(t), \theta_b(t), \mathcal{P}_b(t)\big)$, where $p_b(t)$ denotes the antenna position within the allowable region, $\theta_b(t)$ represents the orientation or beam direction, and $\mathcal{P}_b(t)$ captures the radiation pattern or active configuration. These parameters collectively determine how the antenna interacts with the surrounding propagation environment. This interaction is particularly evident in the wireless channel. In conventional systems with fixed antennas, the channel primarily depends on user location, propagation conditions, and multipath effects. In contrast, fluid antennas introduce an additional dependency on the instantaneous antenna configuration. Accordingly, the channel between user $u$ and base station $b$ at time $t$ can be expressed as $h_{u,b}(t) = f\big(p_b(t), \theta_b(t), \text{location}_u(t), \text{environment}\big)$, where $f(\cdot)$ captures the combined influence of antenna state, user position, and environmental characteristics. This formulation highlights a key distinction: by adapting its antenna configuration, the base station can actively shape the effective channel conditions experienced by its users.

From a network perspective, antenna configuration becomes an additional control variable alongside scheduling, beam selection, and power allocation. These decisions are inherently coupled. For example, repositioning the antenna may improve the signal quality for a target user while simultaneously altering interference patterns across neighboring cells, thereby affecting user association and overall resource efficiency. This coupling leads to a joint control problem in which antenna states and radio resource management decisions must be optimized together.

Solving this joint problem is challenging for several reasons. The antenna configuration space can be large and potentially continuous, particularly when multiple movement and radiation options are available. Network conditions evolve over time due to user mobility, traffic fluctuations, and dynamic interference, while decisions at one base station influence neighboring cells through inter-cell coupling. As a result, fluid antenna networks form high-dimensional and dynamic control systems, where conventional rule-based or static optimization approaches are often insufficient. This complexity motivates the use of learning-based control mechanisms that can adapt antenna configurations and resource allocation policies based on observed network behavior. In the following section, we build on this perspective and introduce a learning-driven control architecture for fluid antenna networks.

\section{AI-Native Control Architecture for Fluid Antenna Networks}

The previous section established that fluid antenna systems introduce an additional control dimension into wireless networks. Once antenna configurations become dynamically adjustable, the network must determine how antenna adaptation should interact with conventional radio resource management decisions. Addressing this challenge requires a control framework that continuously observes network conditions, interprets their implications, and selects appropriate antenna and resource configurations in real time.

To enable this capability, we consider an AI-native control architecture that integrates antenna adaptation into a closed-loop decision process. Rather than treating antenna configuration as an isolated optimization task, the proposed framework embeds antenna control within a broader context of network state awareness and adaptive decision-making.

\subsection{Network State Representation}

Effective control begins with constructing a compact yet expressive representation of the network state. In fluid antenna networks, this state must capture both conventional network information and the current antenna configuration, since antenna positioning directly affects channel conditions and interference patterns. A representative state comprises channel measurements, user locations and traffic demand, base-station antenna configurations, and inter-cell interference indicators. Together, these elements provide a structured description of the network context, enabling the control mechanism to reason about the impact of different configurations on system performance. Importantly, antenna configuration is not an external parameter but an integral component of the network state, as it directly influences the communication environment.

\begin{figure*}[h]
\centering
\includegraphics[width=0.75\linewidth]{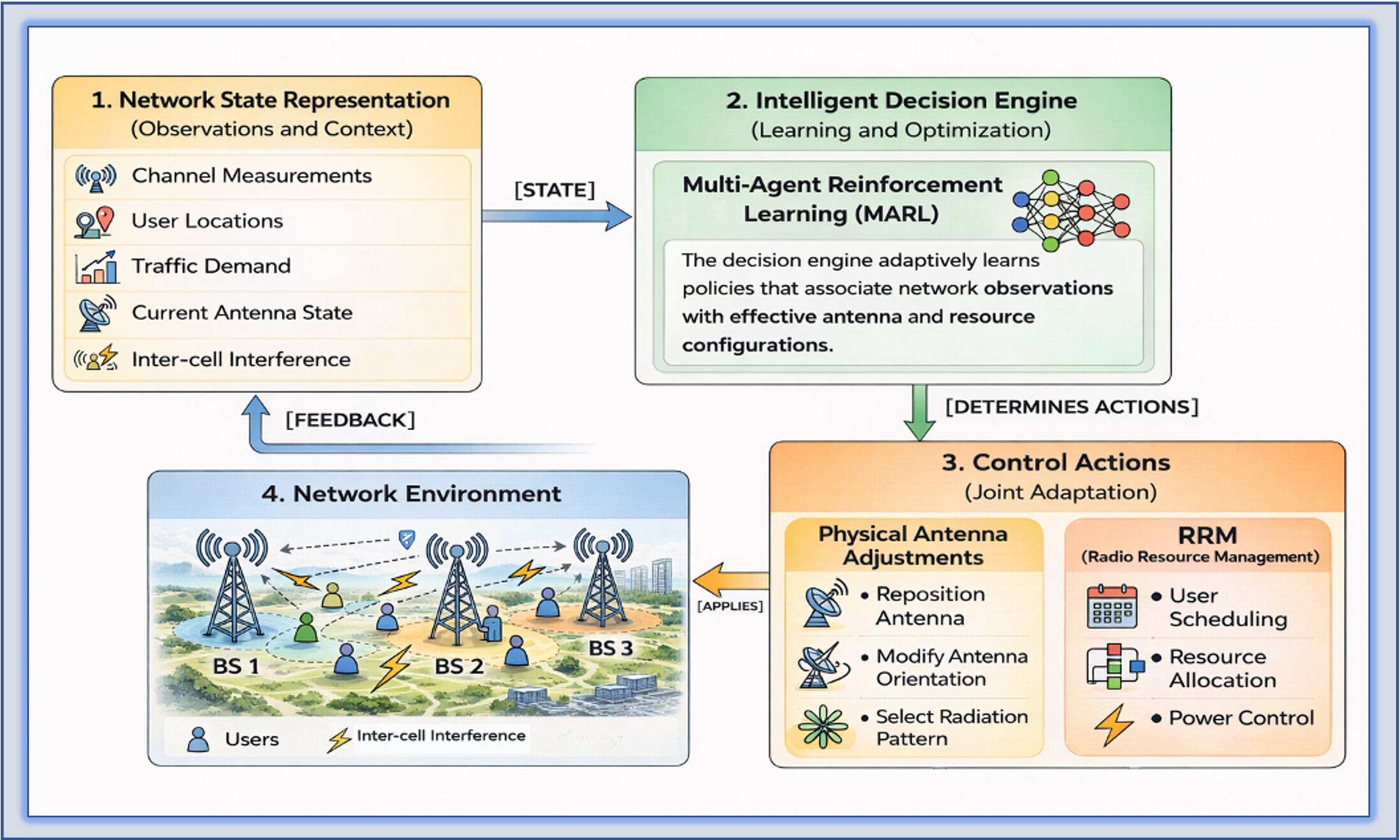}
\caption{AI-native control framework for fluid antenna networks, illustrating how adaptive antenna configurations and radio resource management decisions are jointly optimized through a closed-loop interaction between the network environment and learning-based control mechanisms.}
\label{arch}
\end{figure*}

\subsection{Intelligent Decision Engine}

Given the current network state, the system must determine how to adapt. This functionality is performed by an intelligent decision engine that maps observed states to control actions. Due to the high-dimensional and dynamic nature of fluid antenna networks, conventional rule-based or static optimization methods are insufficient to capture the complex interactions between antenna configurations and network behavior. Learning-based approaches therefore provide a natural solution, with reinforcement learning offering an effective framework for sequential decision-making under dynamic conditions.

By interacting with the network environment and observing performance outcomes, the decision engine can progressively learn policies that associate network states with effective antenna and resource configurations. In multi-cell deployments, this extends naturally to distributed decision-making, where each base station operates as an independent learning agent that adapts to local observations while implicitly coordinating through the shared wireless environment.

\subsection{Control Actions}

Based on the control engine's decisions, the system applies actions that jointly adapt antenna configurations and radio resources. These include both physical antenna adjustments and conventional radio resource management operations.

A base station may reposition its antenna, adjust beamforming directions, or select alternative radiation patterns, while simultaneously determining user scheduling and transmit-power allocation. These actions are inherently coupled: antenna repositioning affects which users can be served efficiently, while scheduling decisions influence the optimal antenna configuration. Treating them independently leads to suboptimal performance, whereas joint adaptation enables more efficient network operation.

\subsection{Closed-Loop Control Model}

The interaction between state, decision, and action forms a closed-loop control system. After actions are applied, the resulting changes in channel conditions, interference patterns, and user experience are reflected in the updated network state, providing feedback for subsequent decisions. This feedback enables the system to continuously refine its behavior and adapt to evolving network conditions, including user mobility, traffic fluctuations, and dynamic interference. Such a closed-loop architecture is particularly important in fluid antenna networks, where the impact of decisions is highly context-dependent. The combination of state-aware decision-making and continuous feedback enables scalable, adaptive control across large, dynamic wireless deployments.

The next section builds on this architecture and illustrates its realization through a multi-agent reinforcement learning framework in multi-cell network scenarios.

\section{Multi-Agent Reinforcement Learning Approach for Fluid Antenna Control}

The control architecture described in the previous section provides a general framework for integrating antenna adaptation into network decision processes. The remaining challenge is how the decision engine can learn to select appropriate antenna and resource configurations under dynamic network conditions. To address this, we consider a multi-agent reinforcement learning (MARL) approach as a representative solution for distributed and adaptive control in fluid antenna networks.

Reinforcement learning is well-suited to such settings, in which decisions are made sequentially under dynamic, partially observable conditions. Each agent observes its local state $s_b(t)$, selects an action $a_b(t)$, and evaluates the outcome through a reward signal $r_b(t)$, where $b$ denotes the base station (agent) index. Over time, this interaction enables the system to learn policies $\pi_b$ that map network conditions to effective control decisions, consistent with the state–decision–action–feedback loop introduced earlier.

\subsection{Learning Framework}

In fluid antenna networks, each base station acts as an independent learning agent $b$ that adapts its antenna configuration and transmission parameters. At each decision step $t$, the agent observes its local state $s_b(t)$, including channel conditions, user demand, interference levels, and antenna configuration. Based on this state, it selects an action $a_b(t)$ according to its policy $\pi_b$, jointly determining antenna repositioning and high-level radio resource allocation. The network then evolves to a new state $s_b(t+1)$ and produces a reward $r_b(t)$ reflecting throughput, fairness, and interference mitigation. Through repeated interaction, each agent refines its policy $\pi_b$ under dynamic conditions.

For intuition, a typical state $s_b(t)$ may include channel quality indicators (e.g., RSRP/SINR), traffic demand, and inter-cell interference levels. An action $a_b(t)$ may correspond to selecting a new antenna position along with scheduling and power allocation decisions. The reward $r_b(t)$ can be defined as a weighted function of user rates and fairness (e.g., proportional fairness), penalized by excessive interference. This formulation enables joint optimization of throughput, fairness, and interference.

\subsection{Distributed Multi-Cell Learning}

In practical deployments, neighboring cells are coupled through interference and shared spectrum, making centralized optimization difficult to scale. MARL provides a scalable alternative, where each base station updates its policy $\pi_b$ based on local transitions $(s_b, a_b, r_b, s_b')$. Although decisions are made locally, agents interact through the shared wireless environment, enabling implicit coordination. For example, improving the signal quality of a local user may alter interference experienced by neighboring cells. Through continuous interaction, agents learn to balance local performance with network-wide interference dynamics, reducing the need for explicit coordination.

\subsection{Learning Workflow and Algorithm}

The learning process follows an iterative closed-loop interaction between agents and the network environment, as summarized in Algorithm~1. At each iteration, each agent observes $s_b(t)$, selects $a_b(t)$ according to $\pi_b$, and receives $r_b(t)$ and $s_b(t+1)$. This feedback captures the impact of antenna and resource decisions on network performance, enabling the agent to evaluate the effectiveness of its actions under evolving network conditions. Over time, agents learn policies that adapt to variations in mobility, traffic, and interference. Learning operates at the network control layer, while beamforming is performed at the physical layer based on the selected antenna configuration, thereby preserving a hierarchical structure. This can be interpreted as a two-timescale process: antenna repositioning adapts to slower channel and interference trends, while beamforming and scheduling operate at faster timescales, allowing the system to balance long-term spatial adaptation with short-term channel tracking.

To illustrate, consider a single step. At time $t$, a base station observes a state in which a cell-edge user experiences low SINR due to strong interference. The agent selects an action that repositions the antenna away from the dominant interference direction while scheduling the user with adjusted transmission parameters. This reduces interference and improves SINR, thereby yielding a higher reward that reflects increased throughput and fairness. Repeating this process enables the agent to progressively learn interference-aware antenna configurations that generalize across similar network conditions.

The next section evaluates the effectiveness of this learning-based control approach through representative simulation results.

\begin{algorithm}[h]
\label{al1}
\small
\caption{MARL for Fluid Antenna Control}
\begin{algorithmic}[1]
\STATE Initialize policy $\pi_b$ for each base station $b$
\STATE Initialize network environment

\FOR{each time step $t$}
    \FOR{each base station $b$}
        \STATE Observe local state $s_b(t)$
        \STATE Select action $a_b(t)$ using policy $\pi_b$
    \ENDFOR

    \STATE Apply joint actions (antenna configuration and high-level resource allocation)
    \STATE Environment evolves (channel and interference update)

    \FOR{each base station $b$}
        \STATE Receive reward $r_b(t)$
        \STATE Observe next state $s_b(t+1)$
        \STATE Update policy $\pi_b$ using $(s_b(t), a_b(t), r_b(t), s_b(t+1))$
    \ENDFOR
\ENDFOR

\end{algorithmic}
\end{algorithm}

\begin{figure*}[h]
\centering
\includegraphics[width=0.68\textwidth]{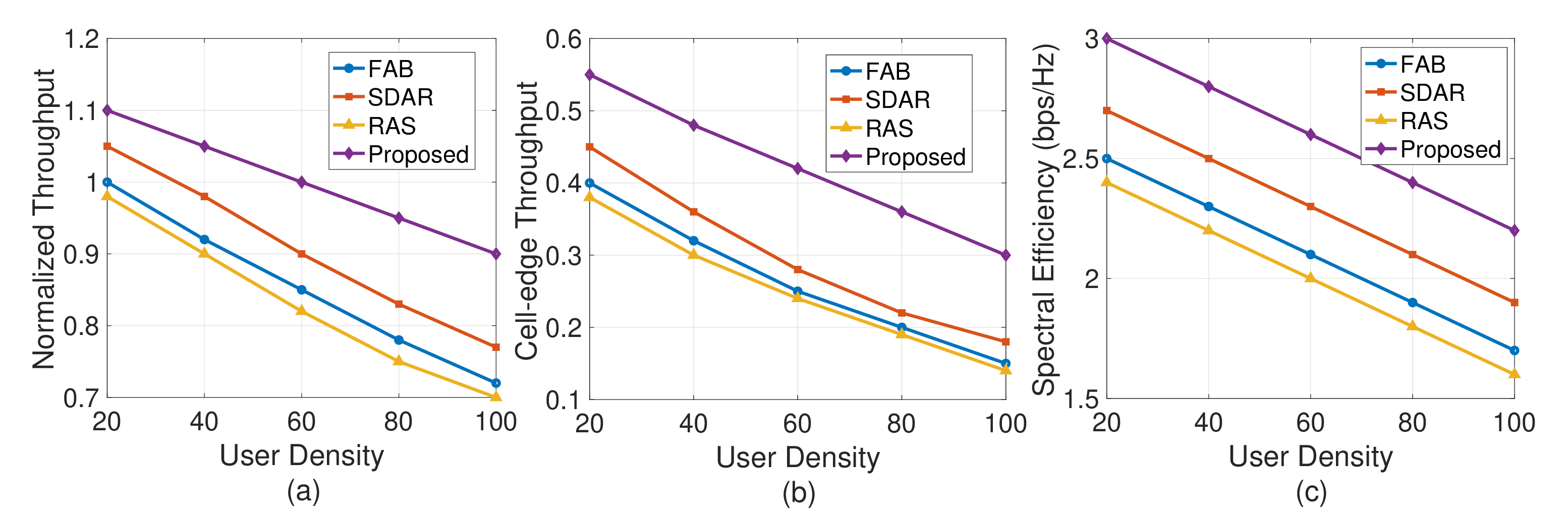}
\caption{Performance comparison under increasing user density. (a) Aggregate network throughput versus user density. (b) Cell-edge throughput (5th percentile user rate), highlighting performance for vulnerable users. (c) Spectral efficiency. The learning-based fluid antenna control yields moderate gains in aggregate throughput but significantly improves cell-edge performance under dense, interference-limited conditions.}
\label{m1}
\end{figure*}

\begin{figure*}[h]
\centering
\includegraphics[width=0.68\textwidth]{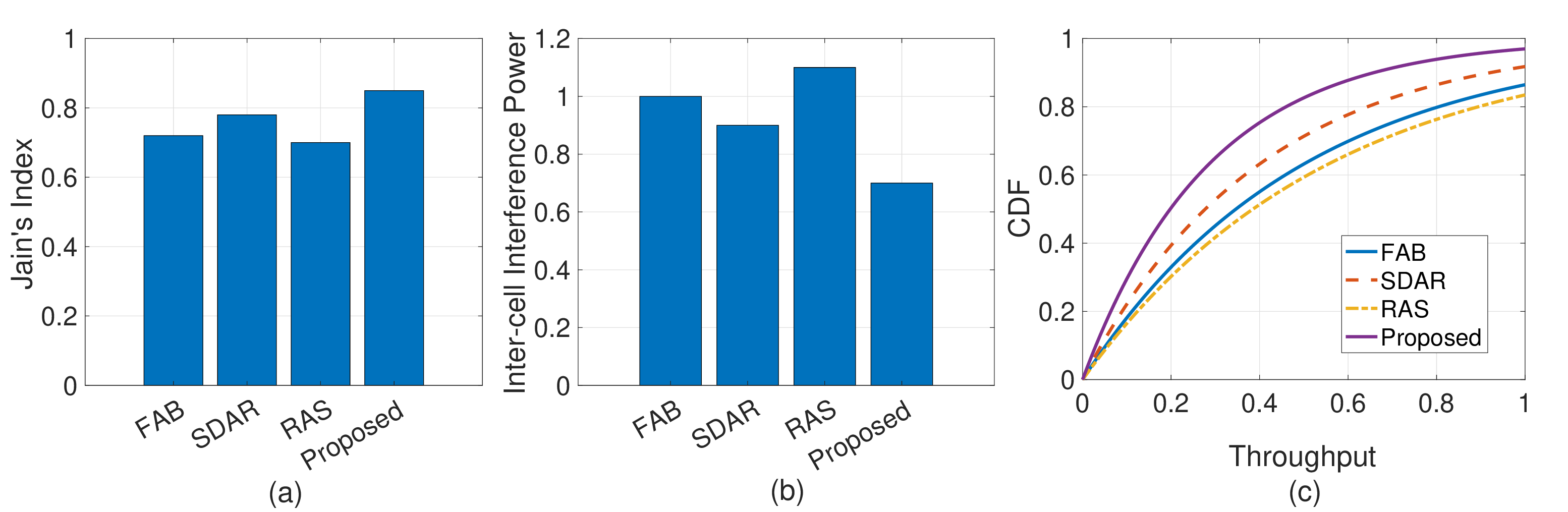}
\caption{User-level and interference-related performance comparison across control strategies. (a) Jain’s fairness index. (b) Average inter-cell interference power. (c) Cumulative distribution function (CDF) of user throughput across the network. The proposed method improves fairness while reducing interference, resulting in a more balanced throughput distribution, particularly benefiting users in the lower-performance regime.}
\label{m2}
\end{figure*}

\section{Illustrative Performance Evaluation}

To illustrate the benefits of intelligent fluid antenna control, we consider a lightweight multi-cell simulation capturing user mobility, inter-cell interference, and antenna reconfiguration. The objective is not exhaustive benchmarking, but to highlight how learning-based antenna control influences performance under representative 6G conditions.

\vspace{-0.2in}
\subsection{Simulation Scenario}

We consider a seven-cell hexagonal network in which each base station employs a fluid antenna that can reposition within a bounded region. The channel model includes large-scale path loss and small-scale fading, while inter-cell interference arises from concurrent transmissions. Users are randomly distributed with moderate mobility, leading to time-varying channel and interference conditions. At each decision interval, base stations adapt antenna configurations and transmission strategies according to the selected control policy. The learning-based controller observes local state information (channel quality, interference, and traffic demand) and selects joint actions comprising antenna positioning and high-level transmission decisions. The control architecture is hierarchical. The learning agent operates at the network control layer, determining antenna configurations and high-level transmission parameters such as user selection and power allocation. Given the selected antenna state, beamforming weights are computed by a conventional physical-layer module. This separation ensures that beamforming operates on top of a dynamically reconfigurable antenna structure, rather than replacing it. In this sense, fluid antenna control complements beamforming by introducing an additional spatial control dimension.

For comparison, we consider three baseline strategies: \textit{Fixed Antenna + Conventional Beamforming (FAB)}, which employs a static antenna with standard beamforming; \textit{Signal-Driven Antenna Repositioning (SDAR)}, which heuristically adapts antenna position based on instantaneous signal strength; and \textit{Randomized Antenna Selection (RAS)}, which selects antenna states randomly without coordination with transmission strategies. Performance is evaluated using aggregate throughput, cell-edge throughput (5th percentile), spectral efficiency, fairness (Jain’s index), inter-cell interference, and user throughput distribution.

\subsection{Performance Results}
Fig.~\ref{m1}a shows aggregate throughput versus user density. The proposed method achieves consistent improvements over baseline approaches, with gains becoming more pronounced in dense scenarios where interference coupling dominates system behavior. More importantly, Fig.~\ref{m1}b highlights cell-edge throughput. The learning-based control provides substantial improvements for users near cell boundaries, where interference is strongest and conventional beamforming alone is insufficient. Notably, the gain at the cell edge is significantly greater than the average improvement, reaching up to 50--70\% higher throughput than baseline strategies under dense user conditions. This indicates that the primary benefit of fluid antenna control lies in protecting vulnerable users rather than only improving average performance. Fig.~\ref{m1}c presents spectral efficiency, showing that joint adaptation of antenna position and transmission strategy enables more effective spatial reuse and interference mitigation. Additional insights are provided in Fig.~\ref{m2}. The proposed approach achieves higher fairness (Fig.~\ref{m2}a) while simultaneously reducing inter-cell interference (Fig.~\ref{m2}b), indicating that performance gains are not obtained at the expense of user imbalance. The throughput CDF in Fig.~\ref{m2}c further confirms that the improvement is distributed across users, with notable gains in the lower tail of the distribution.

The observed gains are primarily driven by the learning agent's ability to avoid interference-sensitive antenna configurations. In dense deployments, the controller learns to reposition the antenna away from spatial regions that create persistent inter-cell interference while maintaining sufficient signal strength for the intended user. For example, in cell-edge conditions, the controller shifts the antenna away from directions that strongly couple with neighboring-cell users, thereby reducing cross-cell interference while preserving the desired link quality. This leads to a significant improvement in effective SINR for edge users, directly translating into the observed gains in cell-edge throughput. Overall, this behavior reflects a dynamic trade-off between interference suppression and signal preservation that cannot be achieved with fixed-antenna beamforming alone.

These results demonstrate that fluid antenna control introduces a new degree of freedom that complements conventional beamforming, enabling more adaptive and interference-aware network operation in dynamic environments.

\definecolor{lightgray1}{gray}{0.95}
\definecolor{lightgray2}{gray}{0.90}
\begin{table*}[t]
\centering
\caption{Practical Challenges and Corresponding Research Directions for Fluid Antenna Networks}
\label{tab:challenges}

\rowcolors{2}{lightgray1}{lightgray2}

\begin{tabular}{p{3.8cm} p{6.5cm} p{6.5cm}}
\hline
\textbf{Challenge} & \textbf{Key Issue} & \textbf{Research Direction} \\
\hline

Control Overhead 
& Frequent antenna updates increase control-plane signaling without proportional performance gains 
& Hierarchical control architectures that decouple fast scheduling decisions from slower antenna adaptation \\

Hardware Reconfiguration Latency 
& Physical antenna movement introduces delays that may exceed channel coherence time 
& Latency-aware control policies aligned with hardware constraints and time-scale separation \\

Multi-Cell Coordination 
& Antenna adaptation decisions are inherently coupled across cells through interference 
& Distributed coordination mechanisms and interference-aware multi-agent learning \\

Energy Consumption 
& Frequent antenna reconfiguration increases the operational energy footprint 
& Energy-aware control strategies that balance performance gains with reconfiguration cost \\

Integration with Future RAN Architectures 
& Antenna control requires tight integration with AI-native and software-defined network frameworks 
& Seamless integration with AI-native RAN architectures through standardized control interfaces \\

\hline
\end{tabular}
\end{table*}

\section{Practical Implementation Challenges}

While FAS introduce new capabilities for wireless networks, bringing these ideas into practical deployments presents several important challenges. Unlike purely algorithmic solutions, fluid antenna control requires tight interaction between hardware reconfiguration, network dynamics, and intelligent decision mechanisms. Understanding and addressing these aspects is essential for integrating fluid antenna systems into future 6G infrastructures.

\subsection{Control Overhead}

A key challenge is the signaling overhead induced by antenna adaptation, which relies on continuous collection of channel measurements, interference indicators, and user demand. Excessively frequent repositioning can burden the control plane without yielding proportional performance gains, especially under slowly varying channels. This introduces a fundamental trade-off between responsiveness and efficiency. Consequently, selecting appropriate control intervals becomes a system-level design problem, motivating hierarchical strategies that decouple fast resource allocation from slower antenna adaptation.

\subsection{Hardware Reconfiguration Latency}

A practical limitation arises from the non-instantaneous reconfiguration of fluid antennas. Mechanical or micro-electromechanical implementations introduce delays between control decisions and their execution, directly impacting the control loop. When adaptation is slower than channel dynamics, performance gains degrade. Therefore, control algorithms must explicitly account for hardware response times and avoid overly aggressive policies, underscoring the need for latency-aware design.

\subsection{Multi-Cell Coordination}

In multi-cell scenarios, antenna configurations are intrinsically coupled through interference. Repositioning that benefits one user may degrade performance in neighboring cells, leading to complex coordination challenges. Fully centralized solutions incur high signaling overhead and require global knowledge, while purely local approaches risk instability and suboptimality. Scalable operation thus necessitates distributed learning and coordination mechanisms for stable fluid antenna control.

\subsection{Energy Consumption}

Dynamic antenna repositioning also introduces additional energy costs. Depending on the implementation, mechanical actuation or electronic switching can lead to non-negligible consumption, particularly under frequent adaptation. Energy-aware control policies are therefore required to ensure that performance improvements justify operational costs, a concern that becomes more pronounced in dense deployments.

\subsection{Integration with Future RAN Architectures}

Fluid antenna control must ultimately align with emerging RAN paradigms. As systems evolve toward software-defined and AI-native architectures, such as open RAN and intelligent controllers, antenna adaptation can be embedded within unified optimization frameworks that jointly address resource allocation, mobility, and interference. Defining appropriate interfaces for seamless interaction with these components is essential for practical deployment.

\section{Open Research Directions}

Fluid antenna networks open a wide range of research opportunities across wireless system design, intelligent control, and adaptive radio hardware. While learning-driven antenna adaptation shows strong potential, several challenges must be addressed before large-scale 6G deployment becomes feasible. This section outlines key directions building on the practical issues discussed earlier.

\subsection{Hierarchical Antenna Control}

A promising direction is hierarchical control, where adaptation operates across multiple timescales. Antenna repositioning typically evolves slower than scheduling or beamforming, enabling a structure that combines fast local decisions with slower network-level coordination. Such designs improve stability and responsiveness while reducing signaling overhead, directly addressing control overhead limitations.

\subsection{Joint Control of Programmable Radio Environments}

Future networks will include multiple programmable elements, such as fluid antennas and reconfigurable intelligent surfaces. Joint optimization of antenna positioning and environmental control significantly expands the design space but also increases coordination complexity. Efficient mechanisms for jointly managing these components remain an open challenge.

\subsection{Learning Under Limited Observability}

Learning-based control depends on accurate observations, yet practical systems operate with partial, noisy, and delayed information. Designing algorithms robust to limited observability is therefore critical. Combining local measurements with prediction or cooperative information sharing offers a promising path forward.

\subsection{Scalable Multi-Agent Coordination}

Large-scale fluid antenna networks require coordination among many distributed agents. Independent decisions may lead to instability or inefficiency due to interference coupling. Scalable solutions based on distributed learning, cooperative policies, and interference-aware coordination are essential for stable operation.

\subsection{Integration with AI-Native Network Architectures}

Fluid antenna control must align with emerging AI-native RAN architectures, where data-driven models and intelligent control operate across layers. This requires standardized interfaces, efficient data pipelines, and flexible frameworks that integrate antenna adaptation with scheduling, mobility management, and orchestration functions. Understanding this interaction remains a key research direction.

The relationship between these challenges and their corresponding research directions is summarized in Table~\ref{tab:challenges}.

\section{Conclusion}

Fluid antenna systems introduce a new degree of freedom by enabling dynamic adaptation of antenna configurations to changing radio conditions. Beyond physical-layer gains, this work highlights their impact at the network level, where antenna configuration becomes part of the control problem.

We presented an AI-native perspective in which antenna adaptation is embedded into a closed-loop framework alongside conventional radio resource management. Within this setting, learning-based approaches, particularly multi-agent reinforcement learning, enable distributed and adaptive decision-making across multi-cell environments. The results show that intelligent antenna repositioning improves throughput, enhances cell-edge performance, and mitigates inter-cell interference compared to static operation.

Looking ahead, fluid antenna networks signal a shift in wireless system design, in which antennas evolve from passive interfaces to controllable network elements. Realizing this vision requires advances in adaptive hardware, scalable learning, and integrated control architectures within AI-native RANs.

Therefore, fluid antennas should be viewed not only as a hardware capability, but as a key enabler of more adaptive, efficient, and intelligent 6G networks.

	\bibliographystyle{IEEEtran}
	\bibliography{ref}

\begin{IEEEbiography}[{\includegraphics[width=1in,height=1.3in,clip,keepaspectratio]{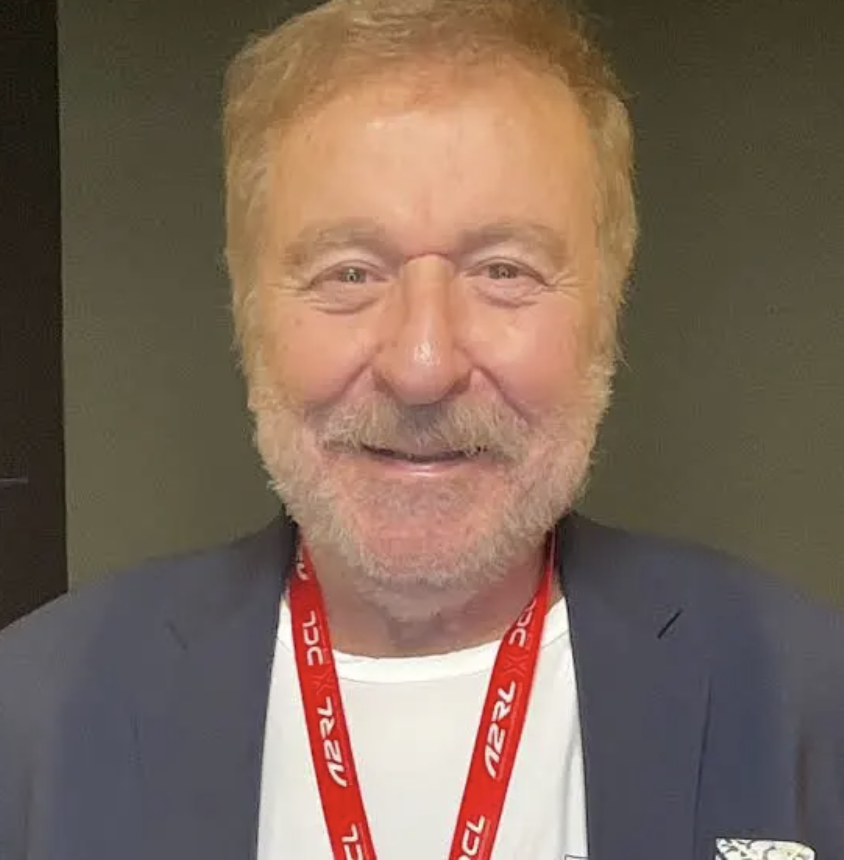}}]{Ian F. Akyildiz}
(Life Fellow, IEEE) received his B.S., M.S., and Ph.D. degrees in electrical and computer engineering from the University of Erlangen–Nürnberg, Germany, in 1978, 1981, and 1984, respectively. From 1985 to 2020, he held the Ken Byers Chair Professorship at Georgia Tech, where he directed the Broadband Wireless Networking Laboratory. A visionary entrepreneur, he is the President of Truva Inc. and a key advisor to global institutions like TII (Abu Dhabi) and Odine Labs (Istanbul). Since 2020, he has served as the founding Editor-in-Chief of the ITU Journal on Future and Evolving Technologies. His pioneering research spans 6G/7G systems, molecular communication, terahertz technology, and underwater networking. As of March 2026, he holds an H-index of 146 with over 155,000 citations. Dr. Akyildiz is an ACM Fellow and recipient of prestigious honors, including the Humboldt (Germany) and TÜBİTAK (Türkiye) Awards.
\end{IEEEbiography}

\begin{IEEEbiography}[{\includegraphics[width=1in,height=1.3in,clip,keepaspectratio]{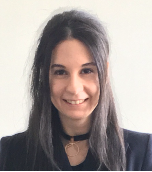}}]{Tuğçe Bilen}
(Member, IEEE) received her B.Sc., M.Sc., and Ph.D. degrees in Computer Engineering from Istanbul Technical University (ITU) in 2015, 2017, and 2022, respectively. She is currently an Assistant Professor in the Department of Artificial Intelligence and Data Engineering at ITU, where she previously served as a Research and Teaching Assistant. Her doctoral research earned several prestigious honors, including the 2025 IEEE Turkey Section Ph.D. Thesis Award, the 2023 Turkish Academy of Sciences (TÜBA) First Prize in Science and Technology, the 2023 Serhat Özyar Young Scientist Honorary Award, and the 2022 ITU Best Ph.D. Thesis Award. Her research focuses on 6G networks, Knowledge-Defined Networking (KDN), AI-driven network management, and digital twins, specifically integrating intelligent systems into future architectures. Dr. Bilen also serves as a reviewer for leading international journals.
\end{IEEEbiography}

\end{document}